\documentclass[draft]{agujournal2019}
\usepackage{url} 
\usepackage{rotating}
\usepackage{multirow}
\usepackage{setspace}
\usepackage{babel}
\usepackage{array}
\usepackage{booktabs}
\usepackage{soul}
\draftfalse

\journalname{Geophysical Research Letters}

\begin{document}
\title{The shape of Jupiter and Saturn based on atmospheric dynamics, radio
occultations and gravity measurements\\}

\authors{E. Galanti\affil{1},Y. Kaspi\affil{1},  and T. Guillot\affil{2}}

\affiliation{1}{Department of Earth and Planetary Sciences, Weizmann Institute of Science, Rehovot, Israel}
\affiliation{2}{Observatoire de la Cote d\textquoteright Azur, Nice, France}

\correspondingauthor{Eli Galanti}{eli.galanti@weizmann.ac.il}

\begin{keypoints}
\item The shapes of Jupiter and Saturn are calculated by jointly fitting their gravity and radio-occultation measurements
\item Saturn's shape has a good match to the radio-occultation measurements, while Jupiter's shape does not
\item The upcoming Juno radio-occultation experiment might give  better constraints on the shape of Jupiter
\end{keypoints}

\begin{abstract}
The shape of the two gas giants, Jupiter and Saturn, is determined primarily by their rotation rate, and interior density distribution. It is also affected by their zonal winds, causing an anomaly of $O(10{\rm km})$ at low latitudes. However, uncertainties in the observed cloud-level wind and the polar radius, translate to an uncertainty in the shape with the same order of magnitude. The Juno (Jupiter) and Cassini (Saturn) missions gave unprecedented accurate gravity measurements, constraining better the uncertainty in the wind structure. Using an accurate shape calculation, and a joint optimization, given both gravity and radio-occultation measurements, we calculate the possible range of dynamical height for both planets. We find that for Saturn there is an excellent match to the radio-occultation measurements, while at Jupiter such a match is not achieved. This may point to deviations from a barotropic flow, something that is to be tested with forthcoming radio-occultation measurements by Juno.
\end{abstract}

\section*{Plain Language Summary}
The shape of the gaseous planets are predominantly set by the rotation rate, but are also affected by the density structure (manifested in the planet's gravity field), and the winds at the planet's outer surface. For both Jupiter and Saturn, the gravity fields have been measured to high accuracy by NASA's Juno and Cassini missions, respectively. This, together with the observed zonal winds, allows an accurate calculation of their shapes. Further constraints can be obtained from radio-occultation measurements, which give radially dependent profiles of density for specific spatial locations. Here we propose a new method for calculating the shape of the gas giants, based on an optimization of the wind latitudinal profile, decay structure, and the polar radius, given both gravity and radio-occultation measurements. We use thermal wind balance to relate the wind to the gravity measurements, and a shape model to relate the wind and polar radius to the radio-occultation measurements. We find that for Saturn there is a good match between the calculated shape and the radio-occultation measurements, while for Jupiter, no such correlation exists. We expect the new radio-occultation measurements to be performed by Juno, to help resolve the shape of Jupiter with a better accuracy.

\section{Introduction}

The gas giants, Jupiter and Saturn, have a shape that is close to an oblate spheroid, set mostly by the solid-body rotation rate, but is also affected by the internal density distribution, and by the zonal winds \cite{Hubbard1982,Kaula1966}. The most prominent characteristic of the shape is a difference of about 10 percent between the equatorial and polar radii \cite{Lindal1981,Lindal1985}. While the shape of a planet with a constant density is an exact ellipsoid, the varying
density inside the gas giants causes deviation in radius in the midlatitudes of around 30~km on Jupiter and 125~km on Saturn \cite{Buccino2020}. The effect of the winds on the shape, named 'dynamical height', is smaller, and was estimated to be around 4~km and 25~km at the equator of Jupiter and Saturn, respectively. However, the uncertainties associated
with the observed cloud-level winds \cite{Garcia-Melendo2011,Tollefson2017}, together with the uncertainty in the polar radius of \textpm 10~km \cite{Lindal1985}, are translated to uncertainty in dynamical height of the same magnitude, rendering the calculation of the dynamical height unattainable \cite{Buccino2020}. In this study we aim to disentangle this uncertainty, taking advantage of the recent gravity measurements and the better understanding the wind profiles.

Knowledge of the exact shape of the gas giants is important for several
research aspects. It is used in determining the position of measurements
by spacecrafts, for example, the Juno's microwave radiometer \cite{Bolton2017,Li2017}.
It can be used to better constrain the rotation rate of the planets
\cite{Helled2009a,Helled2011c,Helled2015}, and has an affect on
the solutions of interior models \cite{Helled2013}. Knowledge of
the shape is also essential in the analysis of radio-occultations
\cite{Lindal1981,Lindal1985}, which in turn are used to validate
the calculated shape.

The calculated shape of the gas giants is commonly compared against
radio-occultation experiments, in which the refraction by the planet's
atmosphere, is measured via a radio signal sent from a spacecraft
to Earth, and then used to calculate the altitude as function of pressure
\cite<e.g.,>{Lindal1981, Lindal1985}. Each radio occultation transect
the atmosphere at a specific geographical point and provides a vertical
profile of density, pressure and temperature, as function of distance
from the planet center. As the measurement depends of the density,
it is by definition a reflection of isopycnal surfaces, and not equipotential
surfaces as in the shape calculation discussed above. Only a few radio
occultations are available for the shape estimation in each planet,
from the Pioneer and Voyager missions (Table~\ref{tab:table1}).
The uncertainty associated with the radio-occultation measurements
is estimated to be around 5~km for both Jupiter and Saturn.
Later missions did not result in additional measurements that 
could be used for the shape estimation. No Jupiter occultations were 
performed by Galileo, and the few published Cassini occultations measurements at 
Saturn did not provide the shape information .  Moreover, due to the growing opacity of the atmosphere
with pressure, most radio-occultation measurements ended at a pressure
of around 100~mb, which is useful for some atmospheric studies, but
less so for interior studies where the 1~bar level is commonly used
as an upper boundary. An estimate of the radii at 1~bar from the
occultations might be obtained by an extrapolation \cite{Helled2009a}. 

Recently, accurate measurement of the gravity field were performed
for both gas giants, with the Juno mission at Jupiter \cite{Iess2018,Durante2020},
and the Grand Finale phase of the Cassini mission at Saturn \cite{Iess2019}.
These measurements enabled the calculation of the wind structure at
both planets, constraining both the possible variations in the cloud-level
wind, and the decay of the flows with depth, setting it to around
3000~km for Jupiter \cite{Kaspi2018}, and around 9000~km for Saturn
\cite{Galanti2019a}. The connection between the cloud-level winds
and the gravity field provides a new constraint that could potentially
improve the calculation of the shape of the giant planets, as the
same winds are also affecting the shape. In addition, the rotation
period of Saturn, once not fully known, with estimates ranging from
of 10~h 32~min 35~s \cite{Anderson2007} to 10~h 39~min 22~s
\cite{Smith1982}, is now calculated from ring seismology to be around
10~h 34~min \cite{Mankovich2019}, and is also supported by \citeA{Read2009}
and \citeA{Helled2015}. These new measurements and studies enable
reexamination of the Jupiter and Saturn shape beyond the approach
taken in \citeA{Buccino2020}. First, the connection between the winds
and the gravity field can potentially reduce the uncertainty from
the the cloud-level wind, which was assumed there to be latitudinally
independent. Second, in that study, the polar radius was not restricted
to modify the dynamical height in unique latitude-independent value,
which led to a large uncertainty envelope around the calculated dynamical
heights \cite[Figure 1]{Buccino2020}. This uncertainty could also
be reduced by looking specifically on how the polar radius affects
the shape derived from the occultation measurements.

In this study, we propose a new method to better constraint the shape
of Jupiter and Saturn, by combining radio occultations and gravity
measurements. We start by calculating the shape using the mean values
of the winds and polar radii. We then explore the range of variations
in the cloud-level wind allowed by the gravity measurements, and calculate
its effect on the dynamical heights. Finally, we use synergistically
the occultations and gravity measurements to find optimal solutions
for the shape for both planets.

\section{Calculating the shape of gas giants\label{sec:shape-calculation}}

The shape of a gaseous planet can be calculated along equipotential
surfaces, based on the measured gravity harmonics, polar (or equatorial)
radius, and wind profile. We follow here the notation of \citeA{Lindal1985}.
First, the gravitational acceleration can be calculated using
\begin{eqnarray*}
g_{r}(r,\phi) & = & -\frac{GM}{r^{2}}\left[1-\sum_{n=2}^{\infty}(n+1)J_{n}\left(\frac{R_{e}}{r}\right)^{n}P_{n}(\sin\phi)\right]+\frac{2}{3}\omega^{2}r\left[1-P_{2}(\sin\phi)\right],\\
g_{\phi}(r,\phi) & = & -\frac{GM}{r^{2}}\sum_{n=2}^{\infty}J_{n}\left(\frac{R_{e}}{r}\right)^{n}\frac{dP_{n}(\sin\phi)}{d\phi}-\frac{1}{3}\omega^{2}r\frac{dP_{2}(\sin\phi)}{d\phi},
\end{eqnarray*}
where ($r,\phi$) are the radius and planetocentric latitude, $G$
is the gravitational constant, $M$ is the planetary mass, $P_{n}$
are the Legendre polynomials, $J_{n}$ are the measured gravity harmonics,
and $R_{e}$ is the reference planetary radius, taken here as the
equatorial radius in both Jupiter and Saturn, as defined in the gravity
measurements \cite{Iess2019,Durante2020} . The rotation rate $\omega$
is composed from the solid body rotation $\omega_{o}$ and the 
contribution from the zonal wind, so that
\[
\omega(r,\phi)=\omega_{o}+\frac{V_\omega}{r\cos\phi},
\]
where  $V_\omega$ is the  cloud-level wind projected barotropically parallel to the planetary spin axis.  The shape of the planet along an equipotential surface can then be
defined with
\[
r(\phi)=R_{p}-\int_{\pi/2}^{\phi}\frac{g_{\phi}}{g_{r}}r(\phi')d\phi',
\]
where $R_{P}$ is the polar radius for a specific pressure, defining
the equipotential along which the shape is calculated. Note that the
planetographic latitude is defined as $\phi+\psi$, where $\psi=\arctan(g_{\phi}/g_{r})$. 

In all previous studies \cite<e.g.,>{Lindal1985,Helled2009a,Buccino2020},
the calculation of the shape was divided into the calculation of the
static body gravitational and centrifugal potential, and then a dynamical
height, resulting from the cloud-level wind, was added to this latitude
dependent shape, thus relaying on a linearization of the problem.
Here, using the derivation discussed above, we perform a direct calculation
that is  potentially more precise, resulting in an improvement compared to previous
studies of up to 1~km in the equatorial region, for both planets.
In order to calculate the effect of the winds on the shape, i.e.,
the dynamical height, we calculate the shape once using the full rotation
rate, $\omega$, and once using only the solid body rotation rate
$\omega_{0}$. The dynamical height is then derived via the subtraction
of the latter from the former.  Note that in both variants of the method there is a fundamental 
assumption that, at the pressure level where the shape is estimated, isopycnals and isobars coincide. 
That is accurate for the part of the shape resulting from the solid body rotation rate, but it also 
implies that the winds there are barotropic (having no change in the direction parallel to the 
planetary spin axis),  at least in the atmospheric regions probed by occultations. In section~\ref{sec:Optimal solutions}  
we will allow for wind variations that might account for baroclinicity between the 100~mb and the cloud levels.

The calculated shape can be compared against the shape derived from
radio-occultation measurements (see discussion above). For most radio
occultations, the deepest pressure for which the radius can be calculated
is around 100~mb.  \citeA{Helled2009a} suggested an extrapolation of the measured radius at  100~mb to a pressure of 1 bar, based on a (latitudinally independent) 48~km height difference along the vertical to the geoid.  We discuss this extrapolation in  Fig.~\ref{fig:basic-shape}. Values of radio-occultation radii, gravity harmonics, and polar radii are given in Table~\ref{tab:table1}.
\begin{table*}
\begin{centering}
\begin{tabular}{>{\raggedright}m{0.5cm}>{\raggedright}p{1.9cm}>{\raggedleft}p{1.2cm}>{\raggedleft}p{1.2cm}>{\raggedleft}p{1.2cm}c>{\raggedright}p{1.8cm}>{\raggedleft}p{1.2cm}>{\raggedleft}p{1.2cm}>{\raggedleft}p{1.2cm}}
 & \multicolumn{4}{l}{\textbf{Jupiter}} &  & \multicolumn{3}{l}{\textbf{Saturn}} & \tabularnewline
\cmidrule{2-5} \cmidrule{3-5} \cmidrule{4-5} \cmidrule{5-5} \cmidrule{7-10} \cmidrule{8-10} \cmidrule{9-10} \cmidrule{10-10} 
 &  & \raggedright{}{\footnotesize{}Measured} & \raggedright{}{\footnotesize{}Gravity optimized} & \raggedright{}{\footnotesize{}Gravity + occultation optimized} &  &  & \raggedright{}{\footnotesize{}Measured} & \raggedright{}{\footnotesize{}Gravity optimized} & \raggedright{}{\footnotesize{}Gravity + occultation optimized}\tabularnewline
\cmidrule{2-5} \cmidrule{3-5} \cmidrule{4-5} \cmidrule{5-5} \cmidrule{7-10} \cmidrule{8-10} \cmidrule{9-10} \cmidrule{10-10} 
\multirow{9}{0.5cm}{\begin{turn}{90}
{\footnotesize{}Gravity harmonics $\times10^{-8}$}
\end{turn}} & {\footnotesize{}$J_{2}$} & \textcolor{gray}{\footnotesize{}-} & {\footnotesize{}48.82} & {\footnotesize{}31.23} &  & {\footnotesize{}$J_{2}$} & \textcolor{gray}{\footnotesize{}-} & {\footnotesize{}-683.06} & {\footnotesize{}-741.75}\tabularnewline
 & {\footnotesize{}$J_{3}$} & {\footnotesize{}-4.50} & {\footnotesize{}-4.51} & {\footnotesize{}-4.50} &  & {\footnotesize{}$J_{3}$} & {\footnotesize{}5.89} & {\footnotesize{}6.74} & {\footnotesize{}3.82}\tabularnewline
 & {\footnotesize{}$J_{4}$} & \textcolor{gray}{\footnotesize{}-} & {\footnotesize{}-4.33} & {\footnotesize{}-4.77} &  & {\footnotesize{}$J_{4}$} & \textcolor{gray}{\footnotesize{}-} & {\footnotesize{}122.71} & {\footnotesize{}113.46}\tabularnewline
 & {\footnotesize{}$J_{5}$} & {\footnotesize{}-7.23} & {\footnotesize{}-7.91} & {\footnotesize{}-7.22} &  & {\footnotesize{}$J_{5}$} & {\footnotesize{}-22.41} & {\footnotesize{}-21.79} & {\footnotesize{}-23.61}\tabularnewline
 & {\footnotesize{}$J_{6}$} & \textcolor{lightgray}{\footnotesize{}1.00} & {\footnotesize{}0.23} & {\footnotesize{}0.99} &  & {\footnotesize{}$J_{6}$} & \textcolor{lightgray}{\footnotesize{}401.44} & {\footnotesize{}397.37} & {\footnotesize{}374.88}\tabularnewline
 & {\footnotesize{}$J_{7}$} & {\footnotesize{}12.00} & {\footnotesize{}11.59} & {\footnotesize{}12.03} &  & {\footnotesize{}$J_{7}$} & {\footnotesize{}10.77} & {\footnotesize{}6.86} & {\footnotesize{}6.13}\tabularnewline
 & {\footnotesize{}$J_{8}$} & \textcolor{lightgray}{\footnotesize{}3.50} & {\footnotesize{}4.41} & {\footnotesize{}3.43} &  & {\footnotesize{}$J_{8}$} & \textcolor{lightgray}{\footnotesize{}-539.77} & {\footnotesize{}-556.97} & {\footnotesize{}-528.23}\tabularnewline
 & {\footnotesize{}$J_{9}$} & {\footnotesize{}-11.30} & {\footnotesize{}-7.88} & {\footnotesize{}-11.17} &  & {\footnotesize{}$J_{9}$} & {\footnotesize{}36.91} & {\footnotesize{}34.00} & {\footnotesize{}32.16}\tabularnewline
 & {\footnotesize{}$J_{10}$} & \textcolor{lightgray}{\footnotesize{}-3.00} & {\footnotesize{}-4.75} & {\footnotesize{}-3.23} &  & {\footnotesize{}$J_{10}$} & \textcolor{lightgray}{\footnotesize{}348.44} & {\footnotesize{}323.44} & {\footnotesize{}308.44}\tabularnewline
 & {\footnotesize{}RMSE of fit} &  & {\footnotesize{}1.13} & {\footnotesize{}0.07} &  & {\footnotesize{}RMSE of fit} &  & {\footnotesize{}7.79} & {\footnotesize{}12.96}\tabularnewline
\cmidrule{2-5} \cmidrule{3-5} \cmidrule{4-5} \cmidrule{5-5} \cmidrule{7-10} \cmidrule{8-10} \cmidrule{9-10} \cmidrule{10-10} 
\multirow{8}{0.5cm}{\begin{turn}{90}
{\small{}100 mb radius (km)}
\end{turn}} & {\footnotesize{}V2N (71.8S)} & {\footnotesize{}67294} & {\footnotesize{}67302} & {\footnotesize{}67298} &  & {\footnotesize{}V1N (71.2S)} & {\footnotesize{}54948} & {\footnotesize{}54943} & {\footnotesize{}54949}\tabularnewline
 & {\footnotesize{}V1N (10.1S)} & {\footnotesize{}71379} & {\footnotesize{}71390} & {\footnotesize{}71384} &  & {\footnotesize{}V2X (26.6S)} & {\footnotesize{}58913} & {\footnotesize{}58914} & {\footnotesize{}58913}\tabularnewline
 & {\footnotesize{}V1X (0.07N)} & {\footnotesize{}71539} & {\footnotesize{}71554} & {\footnotesize{}71548} &  & {\footnotesize{}P11X (9.8S)} & {\footnotesize{}60138} & {\footnotesize{}60144} & {\footnotesize{}60143}\tabularnewline
 & {\footnotesize{}P11X (19.8N)} & {\footnotesize{}70944} & {\footnotesize{}70952} & {\footnotesize{}70946} &  & {\footnotesize{}V1X (2.4S)} & {\footnotesize{}60354} & {\footnotesize{}60356} & {\footnotesize{}60356}\tabularnewline
 & {\footnotesize{}P10X (28N)} & {\footnotesize{}70415} & {\footnotesize{}70413} & {\footnotesize{}70409} &  & {\footnotesize{}V2N (30.5N)} & {\footnotesize{}58545} & {\footnotesize{}58534} & {\footnotesize{}58537}\tabularnewline
 & {\footnotesize{}P10N (60.3N)} & {\footnotesize{}67934} & {\footnotesize{}67931} & {\footnotesize{}67929} &  &  &  &  & \tabularnewline
 & {\footnotesize{}RMSE of fit} &  & {\footnotesize{}9.29} & {\footnotesize{}5.73} &  & {\footnotesize{}RMSE of fit} &  & {\footnotesize{}6.40} & {\footnotesize{}4.63}\tabularnewline
 & {\footnotesize{}$R_{p}$ (km)} & {\footnotesize{}66896} &  & {\footnotesize{}66894} &  & {\footnotesize{}$R_{p}$ (km)} & {\footnotesize{}54438} &  & {\footnotesize{}54437}\tabularnewline
\end{tabular}
\par\end{centering}

  \centering{}\caption{Values for the gravity harmonics, occultations, and polar radii at
100~mb, for the cases shown in Fig.~\ref{fig:basic-shape} and Fig.~\ref{fig:optimal-solution}.
Measured values for the Jupiter (Saturn) gravity harmonics are taken
from \citeA{Durante2020} \cite{Iess2019}, with $J_{6},$ $J_{8},$and $J_{10}$ values shown after subtracting the contribution expected
from the static interior \cite[ for Jupiter and Saturn,  respectively]{Guillot2018,Galanti2019a}.
Shape values from occultations are taken from \citeA{Helled2009a},
with the latitude indicated in brackets. For each planet, shown are
solutions for a gravity only optimization, and a joint gravity and
occultation optimization. Also shown are the RMSE for each type of
measurement.}
\label{tab:table1}
\end{table*}

\section{The shape based on the observed mean values\label{sec:shape-mean-values}}

\begin{figure}[t]
\centering{}\includegraphics[scale=0.4]{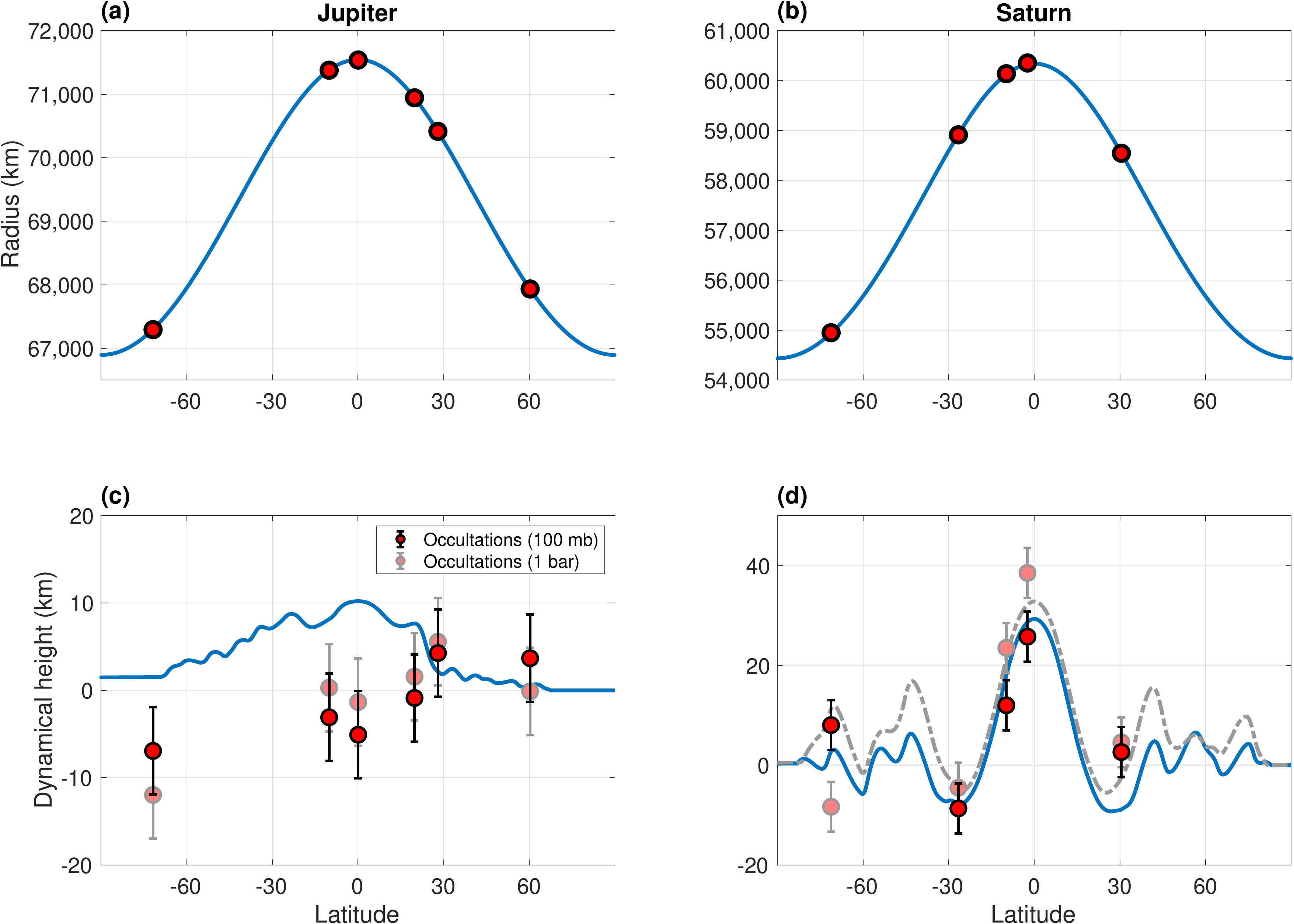}
\caption{(a,b) the full shape of Jupiter and Saturn at 100~mb, calculated
using the mean values of the measured gravity harmonics, polar radius,
and wind profiles. (c,d) the dynamical height contribution
(blue), taken as the difference between the shape with winds and the
shape without winds, and the radii from occultation measurements at
100~mb (red dots), and 1~bar (light-red dots). For the case of Saturn,
shown also is the dynamical height resulting from the gravity-based
wind (dashed-gray).\label{fig:basic-shape}}
\end{figure}

We start with calculating the shape of the planets using the measured
gravity harmonics, observed cloud-level wind, and mean value for the
polar radius (Table~\ref{tab:table1}). The results for the full
shape at 100~mb are shown in Fig.~\ref{fig:basic-shape}, upper
panels. For both planets, the overall shape is consistent with the
occultations (red dots), and is similar to previous results \cite{Lindal1981,Lindal1985}.
Note that the shapes are similar to an ellipse, with differences of
up to $\sim32$~km for Jupiter and $\sim125$~km for Saturn in the
midlatitude regions (not shown). These differences result mostly from
the gravity harmonics $J_{2}$ and $J_{4}$ (see \citeA{Buccino2020} for a detailed analysis).

Next, we compute the dynamical height, calculated via the subtraction
of the shape when calculated with a constant solid body rotation rate,
from the full wind-dependent shape (Fig.~\ref{fig:basic-shape},
lower panels). The radio-occultation measurements at 100~mb are also
plotted with the same solid-body shape subtracted from them as well
(red dots), and we add the extrapolation of the radio occultations
to 1~bar (light red dots). Since we use the same wind profile in
the calculation of both the 1~bar and 100~mbar shapes, the resulting
dynamical height is almost identical (with differences of a few meters).
We therefore present in Fig.~\ref{fig:basic-shape} only the dynamical
height for the 100~mb level. Conversely, the radio occultations differ
between the two pressure levels. This is most likely a result of an
inaccurate extrapolation of the measurements from 100~mbar to 1~bar,
but might also reflect a baroclinic structure at these altitudes.

In the case of Jupiter (Fig.~\ref{fig:basic-shape}c), at both pressure
levels, there is no apparent fit of the wind-induced dynamical height
to that observed by the occultation measurements. The root-mean-square
error (RMSE) is $\sim9$~km at 100~mbar and $\sim8$~km at 1~bar
(Table~\ref{tab:table1}), the same order of magnitude as the dynamical
height itself. It seems that the difference between the dynamical
height and the occultations cannot be removed by modifying the polar
radius, whose uncertainty is about \textpm 10~km, as it will shift
all occultations in the same direction, while the difference from
the dynamical height is negative in the southern hemisphere and positive
in the northern hemisphere. The uncertainty in the cloud-level wind
has the potential to alleviate the difference. We examine these
possibilities in the next sections.

In the case of Saturn (Fig.~\ref{fig:basic-shape}d), we show the
dynamical heights based on two wind profiles, the observed winds (blue)
and the gravity-based optimal wind structure (gray), required to fit
the gravity field \cite{Galanti2019a,Militzer2019}. The fit to the
radio occultations is much better than in the case of Jupiter, with
the dynamical height clearly following the occultation measurements.
Using the optimal wind structure, the fit at the 100~mb level has
an RMSE of $\sim6$~km and at the 1~bar level the RMSE is $\sim4$~km
(Table~\ref{tab:table1}). Therefore, at 100~mb, the RMSE in Saturn
is about 30\% smaller than that in Jupiter, while the magnitude of
the dynamical height is about three times as large. Using the observed
cloud-level wind, leads to a similar RMSE at the 100~mbar, however
it seems that with a constant shift of the occultation values the
fit should be somewhat better. Such a shift can result from a change
in the polar radius. We will examine this possibility in section~\ref{sec:Optimal solutions}.

Note that the results for the dynamical heights for the case of Jupiter
are consistent with those found in \citeA{Buccino2020}. Conversely,
for the case of Saturn, the results are substantially different from
those of the same study, but are quite similar in latitudinal shape
to the profile obtained by \citeA{Anderson2007} with a similar rotation
rate. 

\section{The range of gravity-based wind profiles\label{sec:range-wind-shape}}

\begin{figure}[t]
\centering{}\includegraphics[scale=0.4]{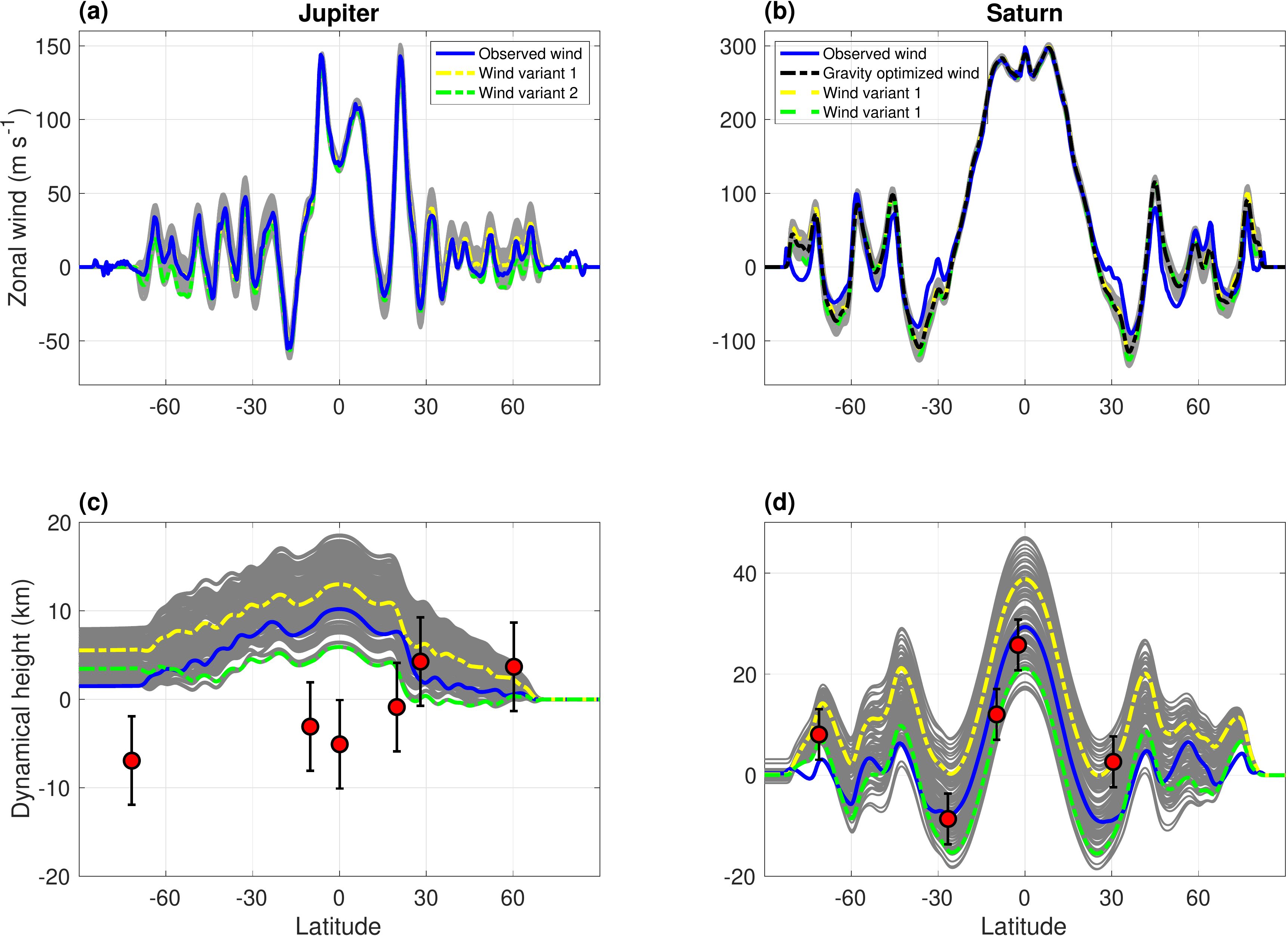}
\caption{(a,b) the range of cloud-level wind (gray) allowing a fit to
the gravity measurements. Highlighted are the observed winds (blue,
\cite{Tollefson2017} for Jupiter, and \cite{Garcia-Melendo2011},
for Saturn) and two example variants (yellow and green).  For the Saturn
case, shown also is the gravity-optimized wind (dashed-black).  (c,d) similar to Fig.~\ref{fig:basic-shape}, but also shown are
the dynamical height resulting from the range of the wind variations
(gray). Highlighted are the dynamical heights for the gravity-optimized
observed wind (blue) and the two examples (yellow and green). \label{fig:wind-range}}
\end{figure}

The results discussed above depend on two parameters that are
not accurately known. First, the polar radius is known to within $\pm10$~km
\cite{Lindal1985}. Modifying the value of the polar radius would
have only a minor effect on the wind-induced dynamical height, but
would shift the radio-occultations values, as they are calculated
with respect to the full shape of the planet. We  discuss this
effect in the next section. Second, the observed cloud-level winds
carry uncertainties from different sources, of up to \textpm 50~m~s$^{-1}$
on Saturn \cite{Garcia-Melendo2011} and up to \textpm 20~m~s$^{-1}$
on Jupiter \cite{Tollefson2017,Fletcher2020}. To address this uncertainty,
we define in both planets an ensemble of 1000 randomly modified cloud-level
winds that are at most \textpm 20~m~s$^{-1}$ (based on the Jupiter wind uncertainty) different from the
observed profiles, and that still ensure the explanation of all relevant
gravity harmonics \cite[see also Supporting Information]{Miguel2022}.
Examples from these ensembles are shown in Fig.~\ref{fig:wind-range}a,b,
together with the observed cloud-level wind of Jupiter and Saturn.
As we show below, for the case of Saturn, uncertainties of \textpm 20~m~s$^{-1}$
are sufficient to account for the differences between the calculated
dynamical height and the radio-occultation measurements.

These wind profiles are then used to calculate the range of dynamical
heights that are consistent with the gravity measurements (Fig.~\ref{fig:wind-range}c,d).
The range of wind profiles translate to a maximal range of $\sim15$~km
for the dynamical height of Jupiter and $\sim30$~km for Saturn (gray
lines). There is no variance in the north pole, as the zonal wind
there is zero by definition, and the maximal variations are reached
around the equator, for which the dynamical height based on the observed
winds is maximal. For the case of Jupiter, it is apparent that the
range of the dynamical height solutions does not help to explain most
of the occultations. Only the occultations at 28$^{{\rm o}}$N and
60.3$^{{\rm o}}$N are within the potential range of the dynamical
heights. Conversely, for the case of Saturn, the range of dynamical
heights contain all occultations.

These results illustrate the potential of fitting the wind-induced
dynamical height with that indicated by the occultation measurements.
While this fit seems attainable for Saturn, for Jupiter there seems
little potential for finding a wind profile that would allow a fit
to the occultations. This seems also the case, when considering the
uncertainty in the polar radius, as it will allow only a constant
shift in the occultations.

\section{Wind profiles fitting the occultation measurements\label{sec:Optimal solutions}}

\begin{figure}[t]
\centering{}\includegraphics[scale=0.4]{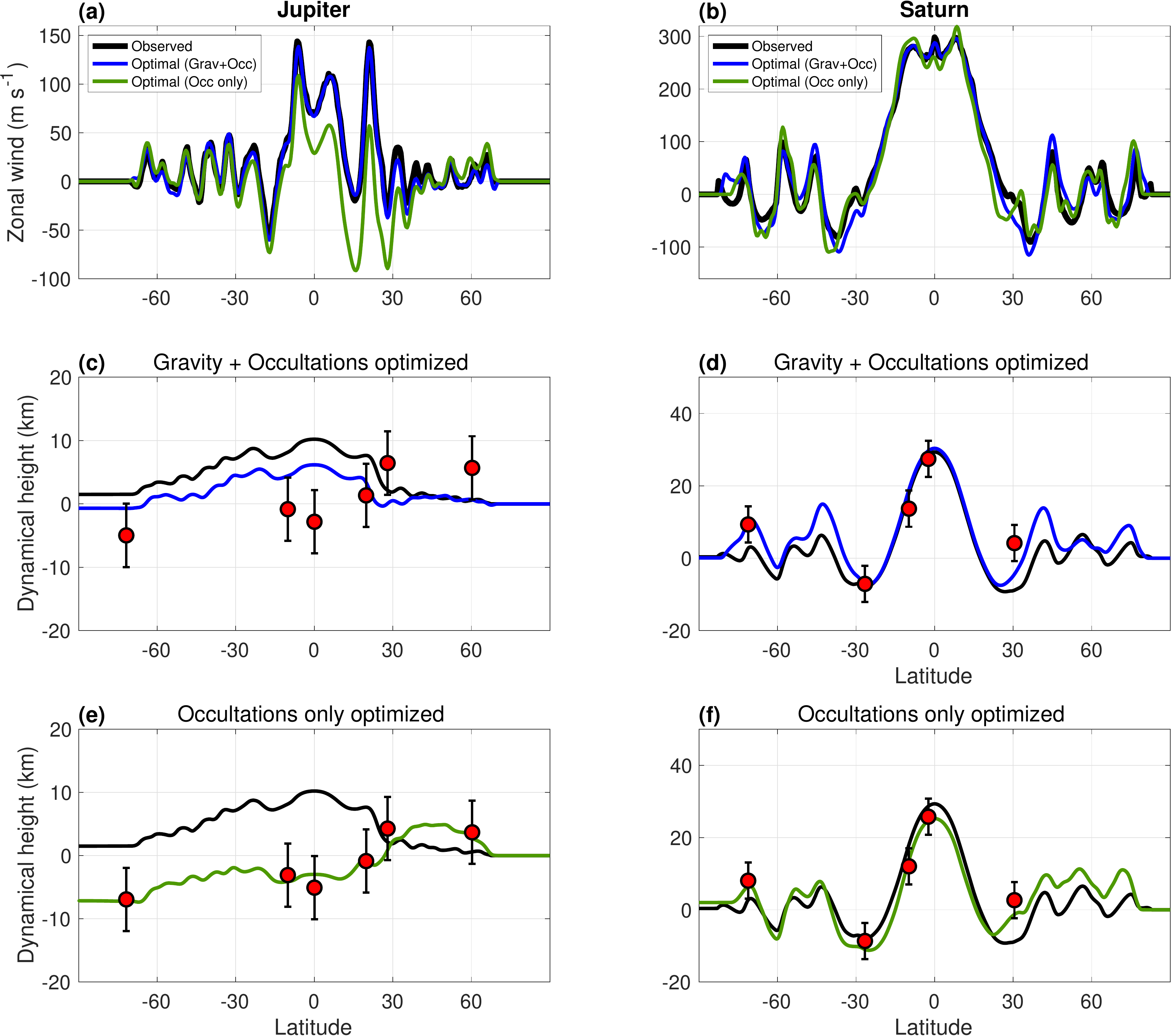}
\caption{(a,b) the cloud-level wind, observed (black), and the optimized
winds, once taking into account both gravity harmonics and occultations
(blue), and the radio occultations only (green). (c,d) dynamical
heights based on the observed wind (black), and the wind optimized
given both gravity and occultation measurements (blue). (e,f) the same as the middle panels, but with the dynamical height based on the optimized wind given the occultation measurements only (green).
Also shown are the occultation measurements (red dots). In the lower
panels, the occultation values are adjusted to the optimized polar
radius (see Table~\ref{tab:table1} for values).\label{fig:optimal-solution}}
\end{figure}

Next, we aim to find specific cloud-level wind profiles, that together
with a specific value for the polar radius, will explain the gravity
and radio-occultation measurements. We start by calculating the optimal
solution for the cloud-level wind profile (latitudinal and depth structure)
and the polar radius, that allow the best fit to both the gravity
field and the occultations. The latitudinal profile of the wind is
allowed to deviate from its observed value, but is kept within the
observed uncertainty of around \textpm 20~m~s$^{-1}$. The polar
radius is allowed to vary within \textpm 10~km around its mean value
(Table~\ref{tab:table1}). The fit to the gravity harmonics and to
the occultations is examined via a cost-function set to give similar
weights to both type of measurements. Finally, the optimal solutions
are searched using an adjoint-based optimization procedure (see Supporting
Information).

For the case of Jupiter, the optimal solution for the cloud-level
wind (Fig.~\ref{fig:optimal-solution}a) is quite similar to the
observed wind. The dynamical height (Fig.~\ref{fig:optimal-solution}c)
is also quite similar, with a smaller amplitude, and the optimal polar
radius is found to be smaller by 2~km than the observed mean value.
The optimal dynamical height has a better fit to the occultations
(Table~\ref{tab:table1}), but the correlation between the latitude
variations of the two data sets is still quite poor. Conversely, for
the case of Saturn, with only minor changes to the gravity cloud-level
wind (Fig.~\ref{fig:optimal-solution}b), the dynamical height matches
all 5 occultations (RMSE of $4.6$~km). This confirms the analysis
of the cloud-level wind range (Fig.~\ref{fig:wind-range}).

The same methodology could be used to find the hypothetical cloud-level
wind profile that allows a full fit to the occultation measurements,
relaxing the restriction on the deviation of the wind from the cloud-level
wind observations, and removing the requirement of fitting the gravity
measurements (see Supporting information). The results are shown in
Fig.~\ref{fig:optimal-solution}e,f) As expected, for the case
of Saturn, with only slight variations of the cloud-level wind, all
radio occultations are perfectly fitted. These variations, when examined
in terms of latitudinal temperature variations due to thermal wind
balance, are of $O(1\,^{\circ}K)$, consistently with observations
\cite{Fletcher2020a}. However, for the case of Jupiter, the change
in the cloud-level wind (Fig.~\ref{fig:optimal-solution}a, green
line) is of the order of 70~m~s$^{-1}$, much more than the observational
uncertainty. Furthermore, the implied latitudinal temperature variations
needed to maintain the difference between this modified wind at 100~mb
and that observed at 1~bar are of $O(10\,^{\circ}K)$,  based on relating the zonal wind shear to the temperature via thermal wind balance,  is much higher than observed at these levels \cite{Fletcher2020a}.

\section{Discussion and conclusion\label{sec:Conclusion}}

Analyses of the shape of Jupiter and Saturn, taking into account recent
understanding gained from the Juno and Cassini gravity measurements,
show very different behavior between the two planets in the dynamical
height exerted by the winds and that observed by radio occultations.  For Saturn, we find an excellent match
between the dynamical height and the occultation measurements at 100~mb.
Optimizing the cloud-level wind and the polar radius, fitting both
gravity and occultations measurements, results in an RMSE of $4.6$~km
in the fit of the dynamical height to the radio-occultation measurements,
less than the uncertainty in the radio occultations. With only minor
modifications of the wind profile, the radio occultation can be perfectly
fitted, when excluding the gravity measurements from the optimization. 

Contrary to Saturn, for Jupiter there seems to be no correlation between
the calculated dynamical height and the radio occultations. No cloud-level
wind profile could be found, with which the occultation measurements
are matched, while satisfying the gravity constraints. Varying the
polar radius does not improve that ability to match the occultations
as well. One reason might be that that the dynamical height in Jupiter
is 3 times as small as that in Saturn, making the uncertainties in
the occultation measurements more prominent. However, the result that
4 out of 6 radio-occultations measurements are completely outside
the calculated dynamical height value (Fig.~\ref{fig:optimal-solution}c),
and have an opposite sign, suggests that something else is causing
the mismatch. Fitting the radio occultation completely requires a
substantially modified cloud-level wind, that is difficult to justify.

The difficulty to match the Jupiter's calculated shape with the radio
occultations can stem from two main sources. First, the radio occultation
radii might be less accurate than the reported uncertainty of \textpm 5~km.
Aside from technical issues that might have occurred during the experiments,
the radio-occultation measurements depend on the knowledge of the
shape itself \cite{Lindal1981}. We now know the shape of Jupiter
to a better accuracy than was estimated in the days of Pioneer and
Voyager, and future reanalysis of these radio occultations might potentially
result in radii that are more consistent with the calculated shape.
Second, it is possible that the atmosphere of Jupiter, in the range
of 100~mb to 1~bar pressure levels, is baroclinic \cite{Kaspi2007},
so that equipotential (pressure) surfaces are not aligned with density
surfaces.  We show that modifying the wind profile to fit
the radio occultations requires  larger latitudinal temperature
variations than observed \cite{Fletcher2021},  but including departures from a barotropic flow 
would also require a modification of the analysis performed here.  Fortunately,  the Juno
 mission is planned to perform a multitude of radio-occultation
measurements in the near future (2023-2024). These measurements, with
their expected wide spatial and temporal coverage, have the potential
to further constrain the shape of Jupiter,  and shed light on the significance of baroclinic effects in its atmosphere.

\section*{Open Research Section}
Data for the calculated shapes and radio-occultations is available via the Harvard Dataverse:
 https://doi.org/10.7910/DVN/B90D83.   The rest of the data used in this study  can be obtained through \citeA{Helled2009a}, \citeA{Iess2018}, \citeA{Guillot2018},  \citeA{Iess2019} and \citeA{Galanti2019a}.

\section*{Acknowledgments}
This study was supported by the Israeli Space Agency and the Helen Kimmel Center for Planetary Science at the Weizmann Institute of Science. The authors thank Prof.  William B.  Hubbard and an anonymous reviewer for the constructive comments.


\end{document}